# Forecasting Bitcoin Volatility: A Comparative Analysis of Volatility Approaches

Cristina Chinazzo* and Vahidin Jeleskovic**

## Abstract

This paper conducts an extensive analysis of Bitcoin return series, with a primary focus on three volatility metrics: historical volatility (calculated as the sample standard deviation), forecasted volatility (derived from GARCH-type models), and implied volatility (computed from the emerging Bitcoin options market). These measures of volatility serve as indicators of market expectations for conditional volatility and are compared to elucidate their differences and similarities. The central finding of this study underscores a notably high expected level of volatility, both on a daily and annual basis, across all the methodologies employed. However, it's crucial to emphasize the potential challenges stemming from suboptimal liquidity in the Bitcoin options market. These liquidity constraints may lead to discrepancies in the computed values of implied volatility, particularly in scenarios involving extreme moneyness or maturity. This analysis provides valuable insights into Bitcoin's volatility landscape, shedding light on the unique characteristics and dynamics of this cryptocurrency within the context of financial markets.

---

* Independent researcher
** Humboldt Universität zu Berlin, email: vahidin.jeleskovic@hu-berlin.de

# 1. Introduction

While Bitcoin reigns as the world's most capitalized and traded cryptocurrency, its status as a financial asset remains a subject of controversy. This controversy stems from the frequent and substantial fluctuations in its value, which understandably concern investors (Yermack, 2015; Bouoiyour & Selmi, 2015).

Recently, numerous research has attempted to determine the fair value and to trace the bubbles in the Bitcoin price series. The previous literature focuses on detecting the level of transaction efficiency in cryptocurrency markets (Cheung *et al*., 2015; Cheah & Fry, 2015; Bartos, 2015; Urquhart, 2016; Kristoufek, 2015; Polasik *et al*., 2015; Bouoiyour *et al*., 2016; Ciaian *et al*., 2016; Balcilar *et al*., 2017).

These prior studies also discuss the reasons behind the volatility of bitcoin. One of the reasons is its spread which is used for speculative purposes and which increases it average volatility while also exhibiting fat tail behavior (Cheah & Fry, 2015; Dyhrberg, 2016; Corbet *et al*., 2018; Hafner; Osterrieder *et al*., 2017; Phillip *et al*., 2018).

Furthermore, news, in particular geopolitical events and governments' willingness to regulate or tax cryptocurrencies have been observed as prominent factors of volatility for bitcoins. One significant recent regulatory development occurred in 2015 when the U.S. Commodity Futures Trading Commission (CFTC) officially categorized Bitcoin and other digital currencies as commodities. It is noteworthy that news events, which often trigger market panic, have a dual impact on volatility. On one hand, the emergence of negative or unexpected news significantly dampens investor expectations, leading to increased selling of the cryptocurrency. Conversely, this selling behavior reduces the overall supply of Bitcoin in circulation, potentially driving up its value due to heightened asset scarcity. Since the volume of transactions is still much smaller than in traditional markets, Bitcoin can be subject to manipulation or influence by large holders who possess huge shares of the total circulating cryptocurrency, Thus, when these big players make any move, it results in severe volatility of cryptocurrencies. Moreover, many authors have noticed that the price of Bitcoin is subject to factors that differ from traditional financial instruments. Kristoufeck (2013) and Glaser *et al*. (2014) found a link between searches on Google and the Bitcoin price trend. Garcia *et al*. (2014) demonstrated that word of mouth on social media and information on Google trends have a substantial influence on the price variations of Bitcoin. Letra (2016) used a forecasting model to study daily quotations and search trends on Google, Wikipedia and Twitter and found that Bitcoin prices are influenced by its popularity. Klein *et al*. (2018) identified a relationship between the volatility in Bitcoin prices and the fear of cyberattacks.

Previous literature predominantly focuses on analyzing Bitcoin price volatility primarily from an investment perspective. This research contributes to and extends the existing body of knowledge by proactively forecasting expected future volatility within the cryptocurrency market. It achieves this by leveraging historical price data and financial derivatives that are currently actively traded.

This study is conducted while considering various existing approaches to calculating volatility: a) Future volatility, which represents the true variability of an asset over a forthcoming period, albeit unobservable and ascertainable only retrospectively. b) Historical volatility, often computed as the standard deviation of past returns, reflecting an asset's price distribution over a prior period. c) Forecasted volatility, entails predicting future volatility over specific timeframes using mathematical methodologies, frequently employing GARCH models (Engle, 1982; Bollerslev, 1986). d) Implied volatility, derived from option prices, signifies the market's anticipation of future volatility through derivative markets (Natenberg, 1994). Given the inherent challenges in directly observing future volatility, this research concentrates on the analysis and comparison of the other three forms of volatility. It aims to assess the associated investment risks in the context of Bitcoin, with a particular emphasis on implied volatility.

Cryptocurrency derivatives markets were first introduced in in 2017. However, after two years the number of transactions has reached a peak (Alexandre, 2019). The reason for investigating derivatives markets for volatility perceived by market players is because of the huge expansion of the Cryptocurrency derivatives markets and total lack of research in this area. The start-up LedgerX is one of the most widespread online platforms (together with Deribit and Quedex) with BTC/USD options quotations (Zaitsev, 2019). This research focuses on this platform, given the sufficiently high number of transactions taking place on it.
– LedgerX was the first Bitcoin option trading platform in the USA, authorized in July 2017 by the CFTC as a Swap Execution Facility and Derivatives Clearing Organization, in order to provide a higher level of transparency, predictability and security for option contracts ("Investment using virtual currency", 2015). The other reason for this choice lies in the fact that the COVID-19 pandemic and the war in Ukraine have led to global economic and energy crises, respectively. Since the beginning of 2020, the global economy has been in an abnormal state. Accordingly, we wish to conduct a study that is representative of the normal state of the global economy and consequently utilize data from before the COVID-19 pandemic.

There is an interesting characteristic associated with implied volatility. Regarding the relationship between option moneyness, which signifies the distance between the current underlying price and the strike price, and implied volatility, it's worth noting that while, for most financial instruments, volatility typically varies depending on the strike price – often with lower values for at-the-money (ATM) options and higher values for options with exercise prices significantly distant from the current market price – this distinct trend in volatility does not appear to be supported by empirical evidence in the context of Bitcoin. Implied volatility tends to remain relatively consistent across various strike prices, suggesting that, despite the typically high fluctuations associated with Bitcoin investments, the volatility component adequately encompasses the inherent risks of the investment. Consequently, additional risks stemming from potential asymmetries in volatility behavior when using derivatives should not be a primary concern (Brown, 2018). However, it's worth noting that some authors argue that a skew profile is beginning to emerge, indicating higher volatility at lower strikes (as seen in "Crypto derivatives," 2018).

The comparison of the above-mentioned three different forms of volatility can help extract the main expectations about the future evolution of the cryptocurrency price according to the market players which is in line with the previous studies (Akgiray, 1989; Latane & Rendleman, 1976; Chiras & Manaster, 1978; Beckers, 1981; Day and Lewis, 1992; Engle & Mustafa, 1992; Bartunek & Chowdhury, 1995).

However, previous studies have primarily applied the mentioned approaches in the context of traditional financial markets. This study contributes to the literature by extending these models and approaches to assess the risk associated with financial investments in cryptocurrencies.

The rest of the paper is organized as follows. Section 2 presents the data along with relevant statistics. We explain the methods in Section 3. Results are presented and discussed in Section 4. The paper concludes with a summary in Section 5.

## 2. Data description

The bid-ask daily prices of the options traded within the timeframe of January 1, 2019, to July 31, 2019, were acquired upon request directly from the online platform LedgerX. LedgerX also furnished the historical series of the underlying BTC/USD, spanning from April 28, 2013, to July 31, 2019 (refer to Fig. 1). It's important to note that both the options and the underlying asset are quoted in US dollars.

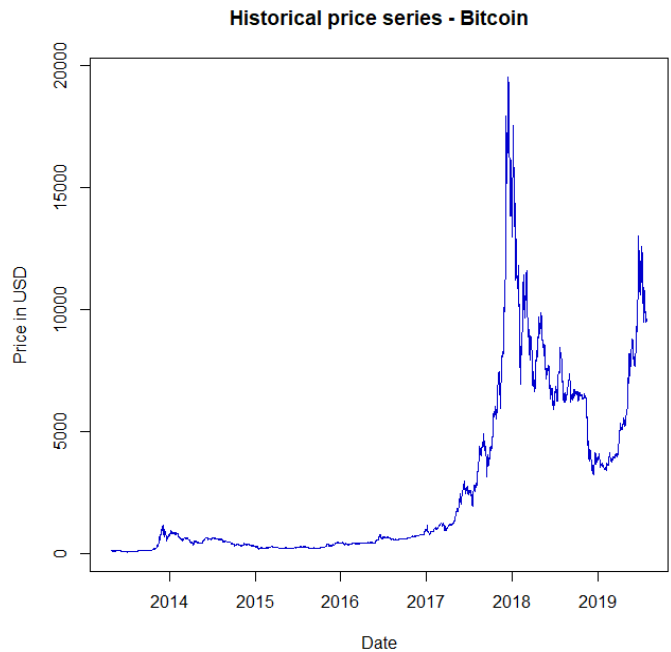

*Figure 1 - Historical Bitcoin prices for the period 28/04/2013-31/07/2019*

Since the cryptocurrency market operates 24/7, the daily data provided in this study corresponds to the prices recorded at 4 p.m. each day. The data exhibited no instances of missing data, eliminating the need for any missing data treatment. Utilizing the historical price series of Bitcoin, we calculated the logarithmic return, as depicted in Figure 2.
The preference for logarithmic return over arithmetic return is well-established, especially when dealing with daily or intra-day observations. Logarithmic return offers the advantage of symmetrically representing price increases and decreases based on the same multiple, differing only by their respective signs. Furthermore, unlike price data, the historical series of logarithmic returns can be considered stationary, as noted by Pichl and Kaizoji (2017). As Figure 2 illustrates, both the returns and the rolling mean and standard deviation exhibit considerable variability over time, indicative of significant fluctuations within the cryptocurrency's historical series. Given the relatively recent emergence of the Bitcoin derivatives market and the substantial variability in the underlying distribution over time, our research did not encompass the entirety of the available price dataset. Instead, we concentrated our analysis on a sufficiently expansive yet closely proximate time frame, specifically spanning from January 1, 2018, to July 31, 2019. Consequently, our calculation of implied volatility exclusively relied on this specific data frame.

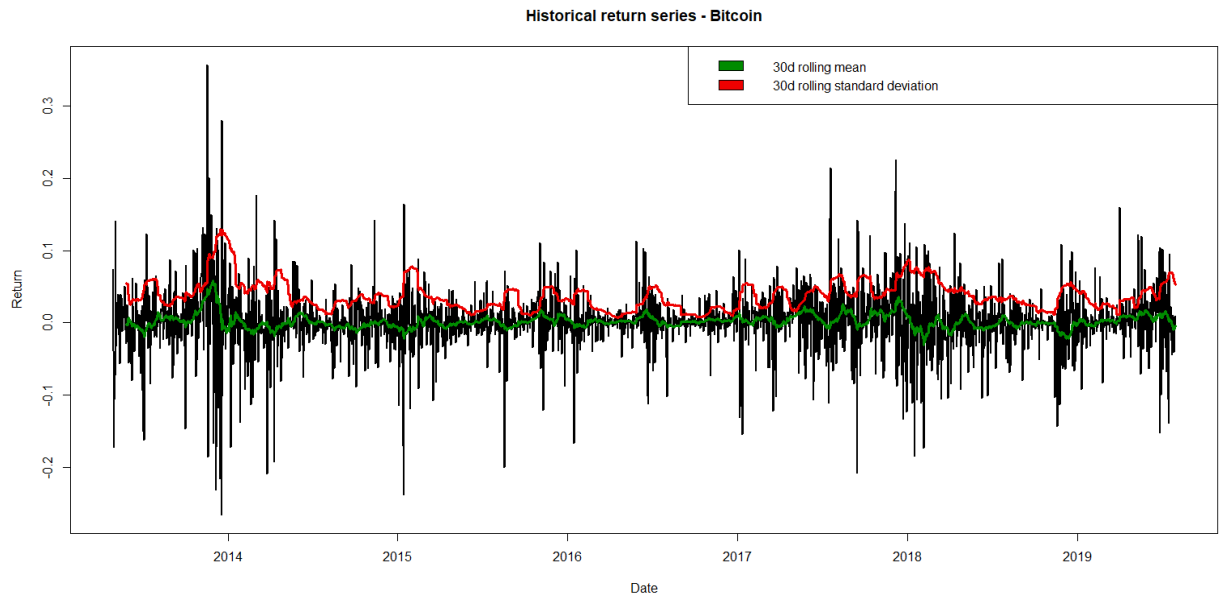

*Figure 2 – Historical series of Bitcoin log returns and rolling moments for the period 28/04/2013-31/07/2019*

When conducting a comparative analysis to identify differences between statistics computed across the entire historical series and those for the limited time spans of 2018-2019 and 2019 (as illustrated in Fig. 3 and Tab. 1), it becomes evident that certain extreme events were excluded from these periods. As a result, several distributional characteristics experienced significant alterations. In an effort to strike a balance between the choice of the time window and its impact on information content, complexity, and redundancy, our research selected the aforementioned one-year-and-seven-month period as appropriate for calibrating volatility forecasting models.

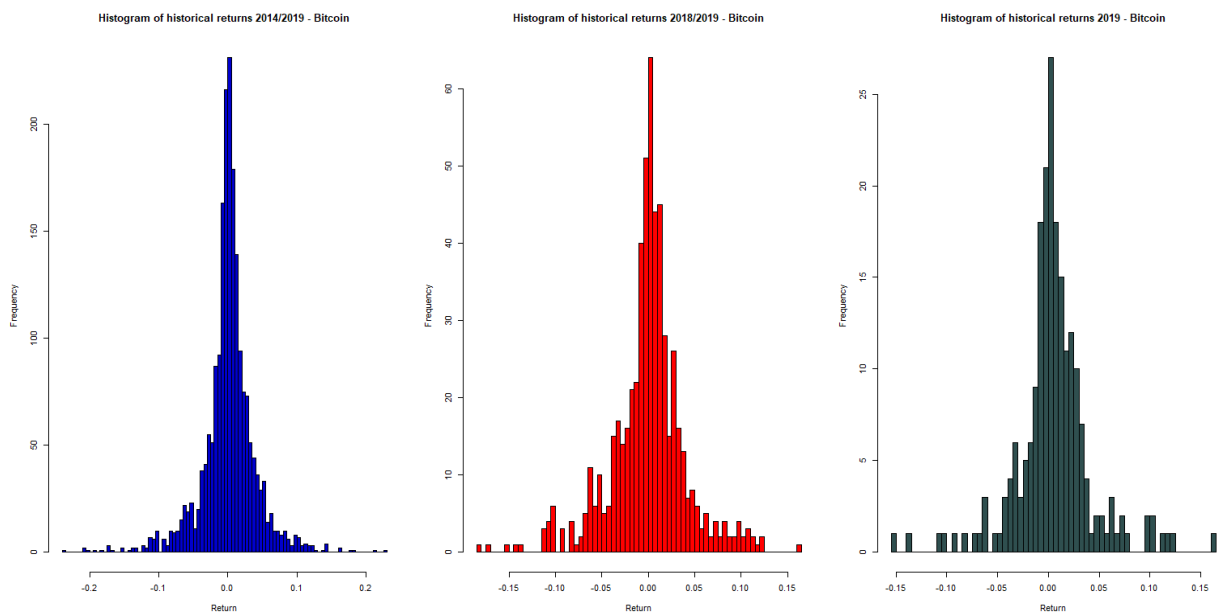

*Figure 3 - Comparison of 6-year, 2-year and 1-year logarithmic return histograms*

This decision aimed to exclude extreme historical events that could have introduced significant distortions in the results for the future period under consideration.

As indicated in Table 1, which provides an overview of the key characteristics of the return distribution for the period 2018-2019, the data exhibit a slight positive skewness and a high kurtosis value. Consequently, it is evident that the Bitcoin series cannot be accurately approximated by a normal distribution, as evidenced by the significantly high value of the Jarque-Bera normality test.

|  | **Period 2014/2019** | **Period 2018/2019** | **Period 2019** |
|---|---|---|---|
| *N. observations* | 2039 | 578 | 213 |
| *Mean* | 0,0013 | -0,0005 | 0,0048 |
| *Median* | 0,0018 | 0,0017 | 0,0025 |
| *St deviation* | 0,0396 | 0,0415 | 0,0386 |
| *Skewness* | -0,3757 | -0,3330 | 0,0193 |
| *Kurtosis* | 8,3713 | 5,6018 | 6,8119 |
| *Min* | -0,2376 | -0,1846 | -0,1518 |
| *Max* | 0,2251 | 0,1600 | 0,1600 |
| *Jarque-Bera stat* | 2506,70 | 176,21 | 133,37 |

*Table 1 - Comparison between 6-year, 2-year and 1-year distribution characteristics*

Regarding the examination of option prices, it is essential to highlight that during our data collection period, LedgerX's platform experienced relatively low liquidity levels, particularly for options with positions that were either significantly in or out of the money. This resulted in notably wide bid-ask spreads for most of the maturities analyzed, as depicted in Figure 4.

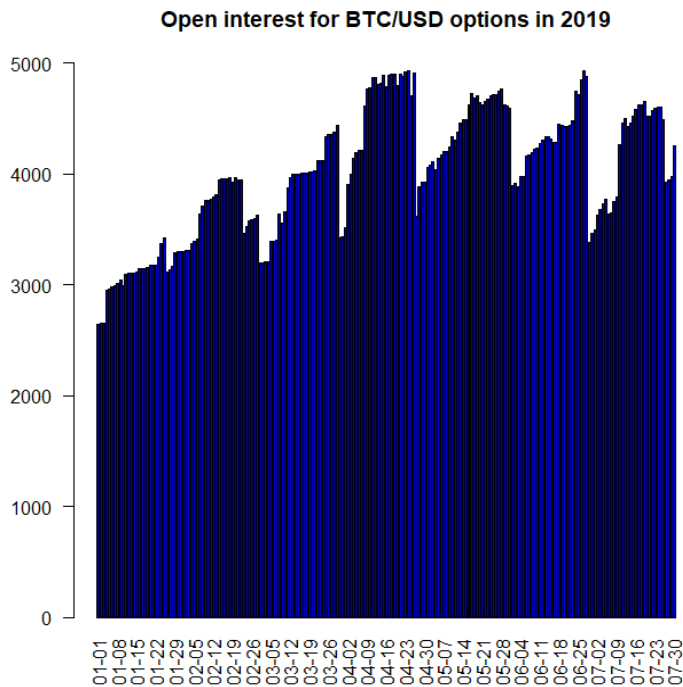

*Figure 4 - Open interests for Bitcoin options for the period 01/01/2019-31/07/2019*

Further investigation into the liquidity status of the market on the final quotation day and its influence on the implied volatility term structure is conducted in Section 3.

Finally, we applied the Augmented Dickey-Fuller test for nonstationarity (Dickey & Fuller, 1979) to the Bitcoin log returns (Fig. 2). The results indicated that nonstationarity must be rejected. Subsequently, we analyzed the heteroscedasticity present in the series using an ARCH test (Engle, 1982) for homoscedasticity, in line with the prevailing literature. Specifically, both forms of ARCH tests, the Portmanteau Q test, and the Lagrange Multiplier test conducted in R, rejected the null hypothesis of homoscedasticity, with p-values close to zero, for both the entire historical Bitcoin series and the 2018-2019 period. We also emphasize that, as determined by a t-test, the means of the Bitcoin log returns considered in this study are not statistically significantly different from zero.

### 3. Methodology

The first form of volatility calculated for the purposes of this analysis is the historical one, understood as simple standard deviation adjusted for the sample (*Eq. 1*).

$$s = \sqrt{\frac{\sum_{i=1}^{N}(x_i - x_{avg})^2}{N-1}} \quad (1)$$

This measure has been calculated on a daily basis using a 30-day rolling window for the period 2018-2019.

For the volatility forecasting, a location scale model for returns' decomposition (*Eq. 2*) has been used: it allows the asset performances to be separated into different components that can be modelled separately in order to obtain a better forecast for the entire return value.

$$R_{t+1} = \sigma_{t+1} z_{t+1} + \mu_{t+1}$$
$$\text{with } z_{t+1} \sim i.i.d. D(0,1) \quad (2)$$

Regarding the component represented by the mean process, especially for short time horizons like daily returns, it is reasonable to assume an average value of returns equal to zero, with the process being primarily influenced by the standard deviation (Christoffersen, 2011). As mentioned earlier and supported by the t-test, we proceed to model the variance without considering the process underlying the mean itself.

The conditional variance fitting procedure has been carried out with a standard GARCH model (Bollerslev, 1986), built considering *p* lags for squared returns – filtered from the mean process, as considered equal to zero – and *q* lags for the variance from the previous periods (*Eq. 3*). It is important to underline that GARCH models detect both the value of conditional volatility – i.e., volatility calculated on the basis of a specific information set, in this case the data from 2018-2019 – and the unconditional volatility, which lets the model consider the fact that the average long-term variance tends to be relatively stable over time and gives an idea of how the variance can move in the long run given the current conditions (*Eq. 4*). The persistence of the model (*Eq. 5*) measures the impact of the current return on unconditional variance for future periods and must be lower than one to guarantee the existence and non-explosivity of the unconditional volatility.

$$\sigma_{t+1}^2 = \omega + \sum_{i=1}^{q} \alpha_i R_{t+1-i}^2 + \sum_{j=1}^{p} \beta_j \sigma_{t+1-j}^2 \quad (3)$$

$$\sigma^2 = \frac{\omega}{1-P} \quad (4)$$

$$P = \sum_{i=1}^{q} \alpha_i + \sum_{j=1}^{p} \beta_j \quad (5)$$

Several extensions of the GARCH model have been introduced later to take into account particular distribution characteristics, which could be relevant in carrying out the analysis of Bitcoin. The integrated GARCH model (Engle & Bollerslev, 1986) has the particularity that $P = 1$, i.e., it assumes that a given shock on the conditional variance is persistent for any future time horizon thus becoming a relevant component in the long run. For this reason, it is particularly useful if the conditional variance is strongly autocorrelated in order to detect the presence of long-term memory in Bitcoin series. In addition, GARCH models capable of weighing positive and negative returns differently have been used in order to capture the leverage effect explained in *Sect. 1*. The Glosten-Jagannathan-Runkle (GJR – Glosten *et al.*, 1993) GARCH introduces in the model an indicator variable $I_t$ that assumes unitary value if the return is negative and zero otherwise; the higher the coefficient $\gamma$ is, the greater the impact of the asymmetrical effect on the final result (*Eq. 6*).

$$\sigma_{t+1}^2 = \omega + \sum_{i=1}^{q} (\alpha_i R_{t+1-i}^2 + \gamma_i I_{t+1-i} R_{t+1-i}^2) + \sum_{j=1}^{p} \beta_j \sigma_{t+1-j}^2 \quad (6)$$

The exponential GARCH model (Nelson, 1991) uses the coefficient $\alpha$ to capture the sign and $\gamma$ to observe the size effect of leverage (*Eq. 7*). This model takes advantage of the logarithmic forms to ensure a positive variance without putting constraints on the values assumed by the coefficients.

$$\ln(\sigma_{t+1}^2) = \omega + \sum_{i=1}^{q}(\alpha_i R_{t+1-i} + \gamma_i[|R_{t+1-i}| - E|R_{t+1-i}|]) + \sum_{j=1}^{p}(\beta_j \ln(\sigma_{t+1-j}^2)) \quad (7)$$

The asymmetric power ARCH model (Ding *et al.*, 1993) is able to take into account both the leverage and the *Taylor effect* (Taylor, 1986) – named after the author who noted that the autocorrelation in absolute returns is usually larger than when computed on squared returns (*Eq. 8*). The term $\delta \in R^+$ represents a Box-Cox transformation of the conditional variance and the coefficient $\gamma$ is expressive of the leverage effect.

$$\sigma_{t+1}^{\delta} = \omega + \sum_{i=1}^{q} \alpha_i(|R_{t+1-i}| - \gamma_i R_{t+1-i})^{\delta} + \sum_{j=1}^{p} \beta_j \sigma_{t+1-j}^{\delta} \quad (8)$$

Finally, consistently with the findings in literature related to the existence of structural breaks and regime changes in Bitcoin series, a Markov-switching GARCH (Haas *et al.*, 2004) has been fitted (*Eq. 9*). It allows the parameters of the model for conditional variance to vary over time according to a latent discrete Markov process ($s_t$ represents a two-states Markov chain depending on a given transition matrix), i.e., it allows to quickly adapt the predictions to changes in the level of unconditional volatility.

$$\sigma_{t+1}^2 = \left(\omega_1 + \sum_{i=1}^{q} \alpha_{1_i} R_{t+1-i}^2 + \sum_{j=1}^{p} \beta_{1_j} \sigma_{t+1-j}^2\right) I(s_t = 1)$$
$$+ \left(\omega_2 + \sum_{i=1}^{q} \alpha_{2_i} R_{t+1-i}^2 + \sum_{j=1}^{p} \beta_{2_j} \sigma_{t+1-j}^2\right) I(s_t = 2) \quad (9)$$

All of the previously mentioned models were computed considering a maximum of 3 lags both for squared returns and conditional variances. For these purposes, the fitting process of the GARCH models (sGARCH, iGARCH, GJR-GARCH, eGARCH) has been done using the *rugarch* package in R (Ghalanos, 2019). The fitting of the Markov-switching models has been carried out using the *MSGARCH* package in R (Ardia et al., 2016).

The goodness of fit of the different models can be verified by computing the respective *log-likelihood* (LL) – calculated on the basis of the probability of observing a given sample production conditioned to the values assumed by the parameters estimated in the model – and by calculating Akaike (AIC, *Eq. 10* – Akaike, 1974) and Bayesian (BIC, *Eq. 11* – Schwarz, 1978) *information criteria* – considered as more consistent than the first method, since they aim to balance the goodness of adaptation of the estimated model and its complexity measured by the number of parameters. Although a *likelihood-ratio test* (*Eq. 12*) is carried out to determine whether the use of an extended model brings significant benefits compared to a simpler one despite the increased complexity, it can be said that the best mody,xclöv-el among the set of plausible models considered is the one which minimizes AIC and BIC-.,,,,,.

$$AIC = -2 \cdot LL + 2 \cdot (p + q) \quad (10)$$

$$BIC = -2 \cdot LL + \ln(n.param) \cdot (p + q) \quad (11)$$

$$LR = 2(LL_{extended\ mod}) - LL_{standard\ mod} \quad (12)$$

It should be noted that in this study, priority is given to the results obtained by minimizing the Bayesian Information Criterion (BIC).

As for the third measure, implied volatility represents the value σ that, *ceteris paribus*, equals the theoretical value given by the Black-Scholes option pricing model (Black & Scholes, 1973) to the real market price of the option (*Eq. 13*). *Eq. 14* shows the Black-Scholes formula for call options, while the model for puts can be derived using the *put-call parity* procedure (*Eq. 15*).

$$C_{mkt} - C_{BSM}(\sigma_{IV}) = 0 \quad (13)$$

$$C_{BSM}(S,t) = S_t N(d_1) - Ke^{-r(T-t)} N(d_2)$$
$$\text{with } d_1 = \frac{\ln\frac{S_t}{K} + (r + \frac{1}{2}\sigma^2)(T-t)}{\sigma\sqrt{T-t}} \quad (14)$$
$$\text{and } d_2 = d_1 - \sigma\sqrt{T-t}$$

$$C_{BSM}(S,t) - P_{BSM}(S,t) = S_t - Ke^{-r(T-t)} \quad (15)$$

The put-call parity formula establishes a unique relationship between the price of a call option and a put option with the same strike price and time to maturity, assuming the absence of arbitrage opportunities (Hull, 2003). The primary assumptions underpinning the Black-Scholes model include: a) the risk-free interest rate remains constant.

b) The statistical process governing logarithmic returns follows a Brownian geometric motion, akin to a random walk with drift, c) both the drift rate and volatility remain constant, d) the underlying asset does not pay dividends, e) there are no arbitrage opportunities in the market, f) borrowing and lending can be done in any amount at the risk-free rate, g) buying and selling any quantity of the underlying asset is possible including short selling, h) the market is frictionless at the option's maturity, with no commissions, taxes, or transaction costs. Despite the stringent nature of these fundamental assumptions in the Black-Scholes model, many derivative platforms, including LedgerX, employ it for pricing options related to cryptocurrencies (Del Castillo, 2017).

Since the real process of the underlying does not threaten the basic intuitions of the model (and its computational simplicity, which make it one of the most widely used pricing models nowadays), the use of the Black-Scholes formula – and the related possibility of extracting the implied volatility from it – has been considered acceptable for this research.

Typically, implied volatility is calculated by market participants starting from the *mid*-price, i.e. the average of the current bid and ask quoted prices, in order to summarize in a single value the expectations of sellers and buyers. However, in the presence of a very wide bid-ask spread, market professionals prefer to calculate two different forms of implied volatility in order to be able to take into account the market liquidity risk ("Should there be two implied volatilities", 2016). Furthermore, due to significantly different prices, calls and puts have also been kept separate in order to analyze them individually and identify any possible intrinsic patterns in implied volatility.

Despite starting from the put-call parity formula implied volatility for calls and puts with same strike and time to maturity should be equal, in reality this does not happen and some graphical differences occur both because of the market liquidity level and the *drift* parameter ($r$ in *Eq. 14*), that – typically positive – inevitably favors with the passage of time the moneyness of call options in comparison to puts.

The typical representation of implied volatility takes the form of a 3D surface that considers – for each option used for the calibration – the strike price, the time to maturity and the related implied volatility. Please note that the market liquidity level is not explicitly represented in it: theoretically, it does not have an effect on the surface, as a high level of liquidity is implicitly assumed. For the following analyses, it is necessary to mention that, in cases of low liquidity such as for Bitcoin, the surface of implied volatility can be distorted or unstable.

A final mention should be made in relation to the aim of observing the variation in implied volatility over time, so that a fourth dimension would be necessary to indicate the passage of the quotation days. Given the impossibility to use this representative form, only the options with an appropriate number of transactions and bid-ask spread throughout the year have been used for this aim, in order to ensure the representation of a volatility surface with a sufficient number of points. In this way, it has been possible to observe the evolution of the implied volatility with the passing of the days in relation to every strike price (fixing therefore the fourth dimension related to the time to maturity).

## 4. Results and discussion

The values of average historical daily volatility in *Tab. 2* are characterized by a great variability during the year, with minimums at the end of 2018 and the beginning of 2019. The daily volatility calculated on 31/07/2019 is rather high (0,0533) and expressive of the presence of large fluctuations in Bitcoin prices in the previous 30 days used for the computation. Projecting this value in the future, assuming a relative stability in volatility and using the *square root of time rule* to scale the computed daily risk factor over longer periods of time, the monthly and annualized values for historical volatility would be 0,2919 and 1,018 respectively. The conversion of daily volatility to a longer time horizon (monthly or yearly) is done by multiplying it by the root of the number of days contained in that frame. These standard deviations are very high and, if verified, would lead to significant fluctuations in the price of Bitcoin (note in fact that the annual value can exceed 100%). It is however important to underline that the rule used for the volatility conversion could theoretically be applied only in the presence of independent and identically distributed returns (Danielsson & Zigrand, 2006). Since this condition is not verified for the Bitcoin series, the calculated estimate may not be consistent with the real volatility extractable from the market for longer time horizons. In any case, the fact that the Black-Scholes model for the calculation of implied volatility also considers the square root of the time to scale the risk factor allows this calculation to be considered significant at least for comparative purposes.

| Date of measurement | Historical volatility |
|---|---|
| 2018-02-01 | 0,0658 |
| 2018-03-01 | 0,0697 |
| 2018-04-01 | 0,0438 |
| 2018-05-01 | 0,0447 |
| 2018-06-01 | 0,0282 |
| 2018-07-01 | 0,0358 |
| 2018-08-01 | 0,0327 |
| 2018-09-01 | 0,0284 |
| 2018-10-01 | 0,0229 |
| 2018-11-01 | 0,0146 |
| 2018-12-01 | 0,0506 |
| 2019-01-01 | 0,0444 |
| 2019-02-01 | 0,0255 |
| 2019-03-01 | 0,0260 |
| 2019-04-01 | 0,0119 |
| 2019-05-01 | 0,0348 |
| 2019-06-01 | 0,0460 |
| 2019-07-01 | 0,0544 |
| 2019-08-01 | 0,0533 |

*Table 2 - Rolling historical daily volatility with a 30-day rolling window for the period 2018-2019*

Regarding the fitting of forecasting models and comparing the results obtained from various models such as standard GARCH, exponential GARCH, GJR-GARCH, asymmetric power ARCH, and integrated GARCH, the information criteria confirm that the standard GARCH (1,1) model performs the best, with AIC and BIC values of -3.66 and -3.64, respectively. However, it's important to note that the differences between these models in terms of information criteria are quite small (Tab. 3).

| Standard GARCH | | | | | | | |
|---|---|---|---|---|---|---|---|
| **AIC** | | | | **BIC** | | | |
| | | q | | | | q | |
| | | 1 | 2 | 3 | | 1 | 2 | 3 |
| **p** | 1 | **-3,6622** | -3,6597 | -3,6528 | 1 | **-3,6396** | -3,6295 | -3,6151 |
| | 2 | -3,6591 | -3,6563 | -3,6563 | 2 | -3,6290 | -3,6186 | -3,6110 |
| | 3 | -3,6563 | -3,6528 | -3,6493 | 3 | -3,6185 | -3,6075 | -3,5965 |
| **Exponential GARCH** | | | | | | | |
| **AIC** | | | | **BIC** | | | |
| | | q | | | | q | |
| | | 1 | 2 | 3 | | 1 | 2 | 3 |
| **p** | 1 | -3,6551 | -3,6516 | -3,6588 | 1 | -3,6249 | -3,6139 | -3,6136 |
| | 2 | -3,6519 | -3,6487 | -3,6604 | 2 | -3,6067 | -3,5959 | -3,6000 |
| | 3 | -3,6559 | -3,6556 | -3,6531 | 3 | -3,5956 | -3,5877 | -3,5777 |
| **Glosten-Jagannathan-Runkle GARCH** | | | | | | | |
| **AIC** | | | | **BIC** | | | |
| | | q | | | | q | |
| | | 1 | 2 | 3 | | 1 | 2 | 3 |
| **p** | 1 | -3,6588 | -3,6563 | -3,6493 | 1 | -3,6286 | -3,6185 | -3,6041 |
| | 2 | -3,5893 | -3,6510 | -3,6496 | 2 | -3,5441 | -3,5982 | -3,5892 |
| | 3 | -3,6482 | -3,6447 | -3,6412 | 3 | -3,5878 | -3,5768 | -3,5657 |
| **Integrated GARCH** | | | | | | | |
| **AIC** | | | | **BIC** | | | |
| | | q | | | | q | |
| | | 1 | 2 | 3 | | 1 | 2 | 3 |
| **p** | 1 | -3,6356 | -3,6351 | -3,6296 | 1 | -3,6206 | -3,6125 | -3,5995 |
| | 2 | -3,6322 | -3,6317 | -3,6294 | 2 | -3,6095 | -3,6015 | -3,5917 |
| | 3 | -3,6313 | -3,6293 | -3,6260 | 3 | -3,6011 | -3,5916 | -3,5807 |
| **Asymmetric power ARCH** | | | | | | | |
| **AIC** | | | | **BIC** | | | |
| | | q | | | | q | |
| | | 1 | 2 | 3 | | 1 | 2 | 3 |
| **p** | 1 | -3,6621 | -3,6609 | -3,6605 | 1 | -3,6243 | -3,6156 | -3,6077 |
| | 2 | -3,6557 | -3,6550 | -3,6577 | 2 | -3,6029 | -3,5947 | -3,5898 |
| | 3 | -3,6487 | -3,6486 | -3,6508 | 3 | -3,5808 | -3,5731 | -3,5678 |

*Table 3 – Information criteria for one-regime GARCH models*

The optimal parameters for the chosen GARCH model (Tab. 4) clearly indicate a significant relationship between the current volatility and its past values (β coefficient), whereas the ARCH component of squared returns (captured by α) plays a relatively marginal role. The high persistence value of 0.9477 suggests the presence of long-term memory or high persistence in Bitcoin volatility, consistent with findings in the existing literature.

| Optimal standard GARCH (1,1) | | |
|---|---|---|
| **Model parameters** | | |
|  | *Coefficient* | *P-value* |
| Omega | 0,0001 | 0,0038 |
| Alpha | 0,0833 | 0,0008 |
| Beta | 0,8644 | 0,0000 |
| **Log-likelihood and information criteria** | | |
| LL | 1.061,39 | |
| AIC | -3,6622 | |
| BIC | -3,6396 | |

*Table 4 - Estimated optimal parameters for GARCH (1,1)*

Based on the estimated parameters for the optimal model, it is possible to forecast the expected daily volatility over a future period of one year. *Fig. 5* shows the estimates made using a *fixed window* perspective, i.e., without updating the model over time and simply proceeding to the volatility forecast over the considered time horizon. For the sake of completeness, it should be noted that a different approach could be used, a *mobile window* estimate, which updates the parameters of the model on the basis of subsequent estimates of daily volatility, so that the most recent forecasts are themselves used for the calibration of the model. It is definite how, in the case of fixed window, the estimates of volatility tend to converge towards the value of long-run average in the long term (this clearly happens in *Fig. 5*).

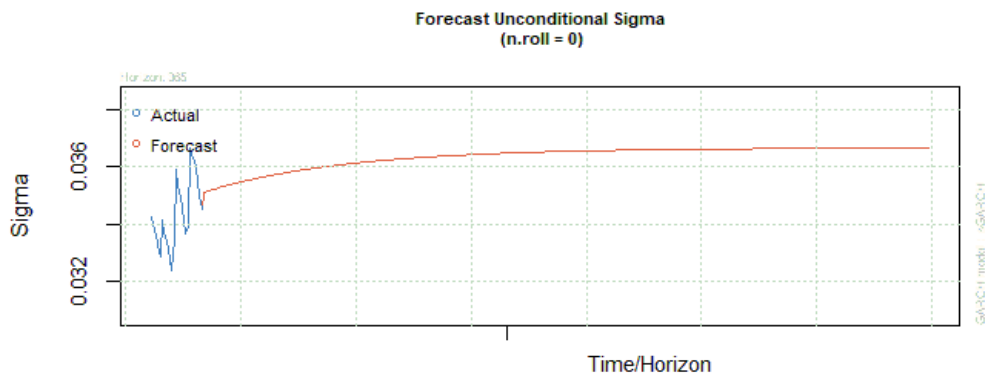

*Figure 5 – 1-year daily conditional volatility forecast with GARCH (1,1)*
However, this does not happen in the case of mobile window but the disadvantage here comes from a strong influence from no longer real but only estimated values for the long-term estimates. For both of the two cases, the reliability of the forecasts is, thus, higher in the short term.

Even though the standard GARCH (1,1) represents the optimal model on the basis of the information criteria and all the models tend to confirm the predominance of the parameter $\beta$ over $\alpha$, it is interesting to analyze the leverage effect results obtained for the optimal order of the other GARCH models (*Tab. 5*).

| **Exponential GARCH (1,1)** | | |
|---|---|---|
| | *Coefficient* | *P-value* |
| Omega | -0,5676 | 0,0000 |
| Alpha | 0,0025 | 0,9208 |
| Beta | 0,9094 | 0,0000 |
| Gamma | 0,1995 | 0,0000 |
| **GJR-GARCH (1,1)** | | |
| | *Coefficient* | *P-value* |
| Omega | 0,0001 | 0,1224 |
| Alpha | 0,0849 | 0,0733 |
| Beta | 0,8643 | 0,0000 |
| Gamma | -0,0026 | 0,9529 |
| **Asymmetric power ARCH (1,1)** | | |
| | *Coefficient* | *P-value* |
| Omega | 0,0000 | 0,9418 |
| Alpha | 0,0129 | 0,0570 |
| Beta | 0,9473 | 0,0000 |
| Gamma | 0,0350 | 0,6018 |
| Delta | 3,4964 | 0,0000 |

*Table 5 – Estimated optimal parameters for the GARCH models when modelling leverage effect*

As for the eGARCH model, the coefficient $\gamma$ – which expresses the size of leverage effect – is positive and quite large, contrary to what it is expected to be in the presence of a typical asymmetric effect (Chang & McAleer, 2017): its positivity implies *ceteris paribus* an increase in volatility when the squared return is higher than its expected value. Consistent with the current literature, one can therefore confirm an inverse leverage effect for the Bitcoin process. However, the GJR-GARCH model does not seem to confirm the leverage effect found in the previous model: the $\gamma$ value is in fact negative and very small and has an insignificant p-value. Finally, the apARCH model also shows a positive value for $\gamma$ which is not statistically significant. Therefore, in this research, it is not possible to demonstrate the existence of a measurable leverage effect in the Bitcoin series through the GARCH models with certainty.

These models essentially consider two regimes in conditional volatility, particularly in terms of the leverage effect. However, the Markov-switching GARCH model (MS-GARCH) offers more flexibility in modeling different states of volatility, and it is not limited to just two regimes. Nonetheless, for the purposes of our analysis, we proceed with the assumption of two regimes as the standard specification when applying the MS-GARCH model, which provides a more intuitive interpretation. The optimal MS-GARCH model, derived through an iterative optimization process, represents a dual regime characterized by two standard GARCH sub-models, as shown in Table 6, which presents the parameters and likelihood measures. It's worth noting that we observed only weak significance in the coefficient estimates for this MS-GARCH model.

| Optimal Markov-switching GARCH | | |
|---|---|---|
| | Standard GARCH (1) | Standard GARCH (2) |
| Omega | 0,0000 | 0,0002 |
| Alpha | 0,0334 | 0,0528 |
| Beta | 0,9084 | 0,9319 |
| P | 0,7721 | 0,5804 |
| **Transition matrix** | | |
| | t+1\|k=1 | t+1\|k=2 |
| t\|k=1 | 0,7721 | 0,2279 |
| t\|k=2 | 0,5804 | 0,4196 |
| **Stable probabilities** | | |
| State 1 | | 0,7181 |
| State 2 | | 0,2819 |
| **Log-likelihood and information criteria** | | |
| LL | | 1.146,9371 |
| AIC | | - 2.277,8742 |
| BIC | | - 2.242,9976 |

*Table 6 – Estimated optimal parameters for the optimal MS-GARCH model*

When analyzing the estimated coefficients, it can be noticed that the two sub-models have a clear predominance of the $\beta$ parameter compared to $\alpha$, even if the second state takes into consideration a higher level of persistence. It is interesting to underline how – even in the presence of multiple regimes – the optimal framework to model the volatility of returns seems to be a standard GARCH model, testifying that there is no need to use more complex extensions. The coefficients of the two sub-models are close to those computed for GARCH (1,1) in *Tab. 4*, but it should be noted that in both of them $\alpha$ is lower and $\beta$ is higher, further reinforcing the already very high influence of the past volatility values in determining the estimated volatility for the future.

When forecasting daily volatility for the upcoming year using the calibrated Markov-switching model (Fig. 6), it becomes evident that the conditional volatility tends to converge towards its unconditional value, which is close to the value predicted by the optimal GARCH (1,1) model. However, the important advantage of the MS-GARCH model lies in its flexibility for forecasting over longer time horizons, as depicted in the comparison between Fig. 5 for GARCH and Fig. 6 for MS-GARCH.

For the sake of comparison, it's worth noting that the predicted long-term value for daily fluctuations is lower than the one estimated using historical volatility, which is strongly tied to its last computation on July 31, 2019.

Regarding the approach for the analysis of implied volatility, as anticipated in *Sect. 3*, bid and ask prices are kept separate to form two different charts, given the low liquidity of the market and the size of the spread.

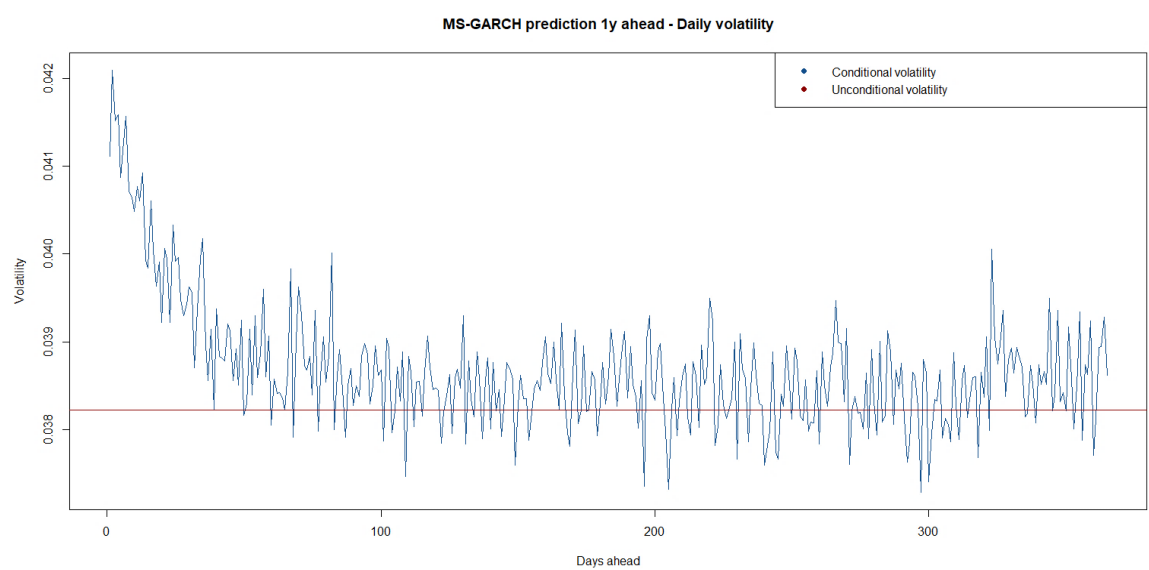

*Figure 6 – 1-year daily conditional volatility forecast with MS-GARCH*

Fixing the observation date to the last available day of quotation (31/07/2019), the structure of the volatility can be seen in in 2D (strike price on x-axis and implied volatility on y-axis) representing the different maturities as level curves so as to make them more easily visually comparable (*Fig. 7*). As can be seen from the charts, the options quoted on LedgerX cover a very wide range of strike prices – even reaching the level of $100.000, not shown since it has an open interest close to zero due to the extreme moneyness – and expiration dates. Given the value of the underlying equal to $9.767,60 on 31/07/2019 and although the trend of the curves is unstable and not clearly-defined because of the low liquidity, higher values for underlying prices very distant from the actual level and lower volatility in correspondence of approximately ATM options can be noticed as almost all the volatility graphs tend to assume the typical smiling shape. The curves related to options with maturities of less than one month show a very unstable and anomalous trend, confirming that – as described in *Sect. 1* – the calculation of volatility on very short-term options is strongly discouraged given the strong impact of any new information and the absence of residual time to dampen its effects.

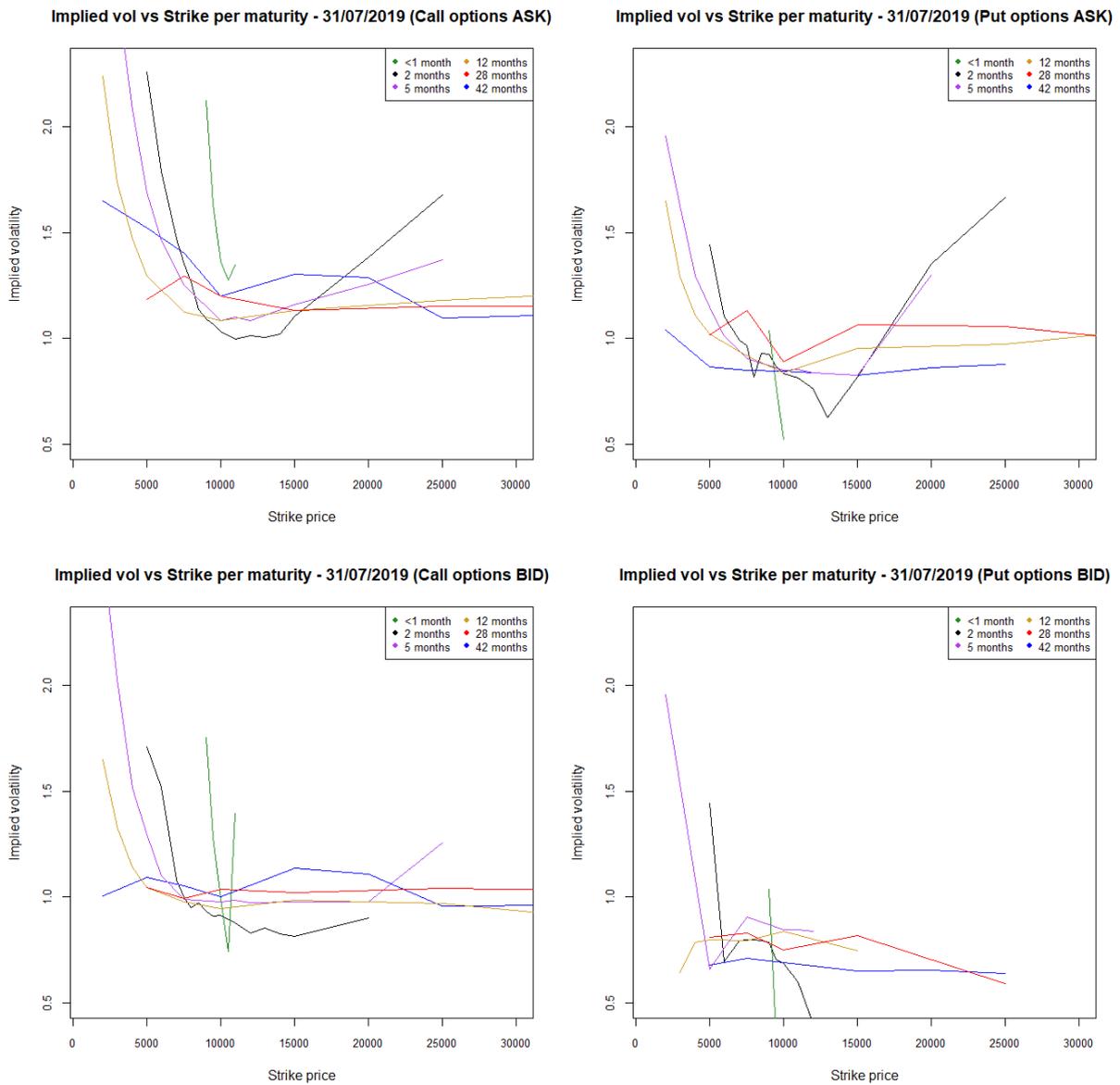

*Figure 7 - Implied volatility smile per time to maturity on 31/07/2019 (separate for call/put and bid/ask prices)*

Similarly, options with very long maturities have an unusual pattern compared to others, resembling flat curves that do not substantially differentiate according to the strike price. This trend, contrary to very short-term options, is related to the fact that the market tends to not incorporate highly sensitive information into these prices, assuming that it can automatically reabsorb over time.

As for the shape of the smiles, it must be said that they tend to a skew: for very high strike prices – for which, given the same underlying level, calls are OTM and puts are ITM – the volatility tends to be significantly lower than for options with smaller strikes. One of the explanations underlying the formation of the skewed shape is related to the log-normal returns assumption of Black-Scholes model, which in reality often does not occur; indeed, the underlying distributions for Bitcoin are characterized by fat tails and extreme yields. Therefore, the operators in Bitcoin market tend to take advantage of put options OTM ( small strikes and typically a low price) to protect themselves from the possible downward price movements in the underlying or, possibly, to assume short positions on the cryptocurrency. Contrary to the findings in previous literature, which did not establish a link between volatility and strike price, the observed trend in Bitcoin's case resembles what is typically observed in equity markets. When share prices decline, there is greater uncertainty, resulting in an expected increase in the effective volatility of the underlying asset. As share prices rise, the level of certainty increases, leading to more orderly price movements and a decrease in volatility.

It is important to note that some implied volatility values in the charts exceed 100%, indicating market expectations for significant future price movements in Bitcoin. This phenomenon is particularly expected for options with distant strike prices since reaching these price levels within a relatively short period would require anticipating substantial price swings in the underlying asset.

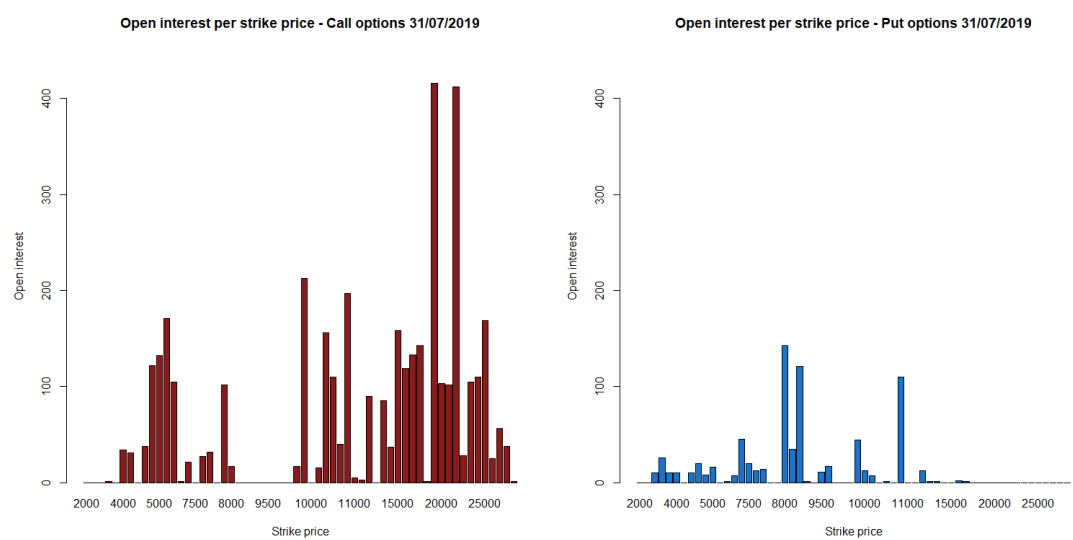

*Figure 8 - Open interest for call and put contracts per strike price on 31/07/2019*

Additionally, it is worth highlighting that, for certain listed options, it was impossible to calculate implied volatility using the Black-Scholes formula due to the iterative procedure's failure to find a volatility value that converges with the theoretical market price of the option. This issue is likely attributed to the low level of liquidity in the market, which is evident from the limited number of open contracts, especially for options with extreme strike values. As such, the obtained values of implied volatility should be interpreted with caution.

The high bid-ask spreads observed for each of the options traded on July 31, 2019, further corroborate the limited liquidity in the market (Fig. 9).

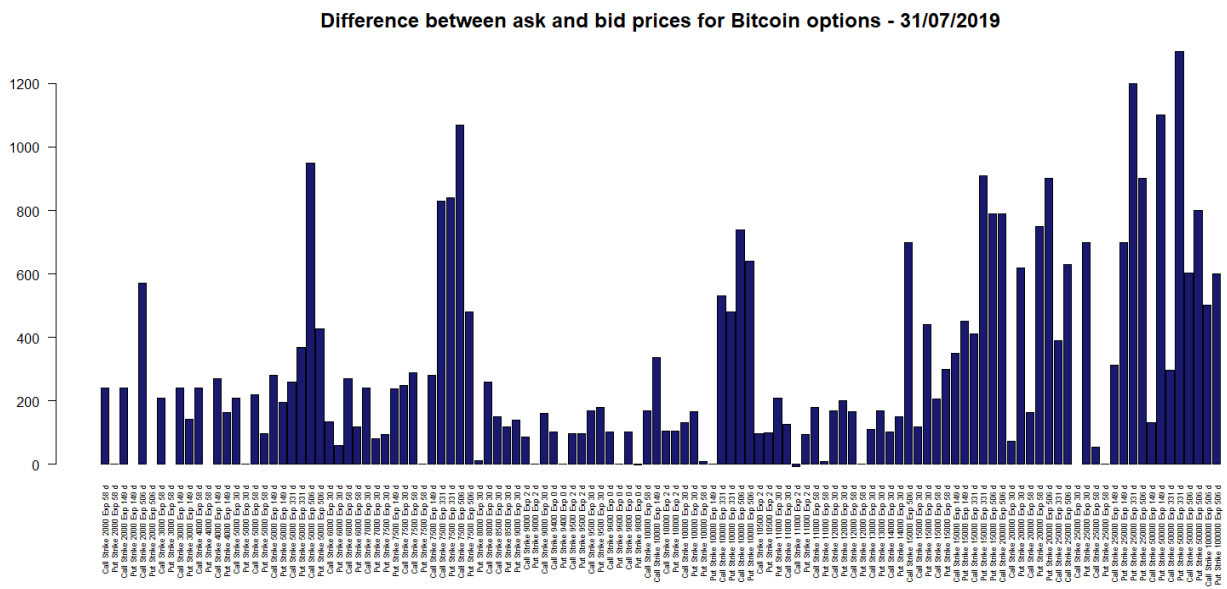

*Figure 9 - Bid-ask spread by option type on 31/07/2019*

In conclusion, to analyze the evolution of implied volatility over the last year of trading and the changes in the market over time, we need to fix the option's time to maturity. This is primarily due to the low liquidity of some options with very short or very long maturities and the practical challenge of incorporating an additional dimension (beyond implied volatility, strike price, and time of the year) into the graphical representation. Therefore, we focus on options with a medium-term maturity, approximately six months, as they exhibit higher liquidity and a more distinct and regular skew compared to other maturities, as shown in Fig. 7.

To create a stable and comprehensive view of the market's situation throughout the year, we average the implied volatility values for both bid and ask prices for calls and puts with the same strike price and maturity. This approach smooths out the results obtained for individual cases, resulting in a volatility surface that incorporates all available data. Additionally, we summarize only the primary strike prices, rounding them to the nearest thousand, to avoid anomalous values caused by low open interests.

The evolution of implied volatility over time is represented by dividing the observations into 38 sub-periods, and the volatility values are averaged within each sub-period. This approach accounts for the inability to calculate likely volatility values for certain options with extreme moneyness and the presence of substantial missing data on specific trading days, as mentioned in Section 2.*Fig. 10* shows, as could be expected given the low liquidity level of the market, a surface with a lot of missing data, corresponding to non-quoted strikes in that given moment. At the beginning of 2019 the quotations were clearly concentrated on rather low strikes, avoiding all the values above 8.000. During the last months, the prices are instead concentrated on the highest strikes creating a gap in correspondence of very low values. Therefore, a shift in expectations during the year can be observed: the market tends to move towards higher strike prices, testifying to a rise in the expected Bitcoin price compared to the previous months.

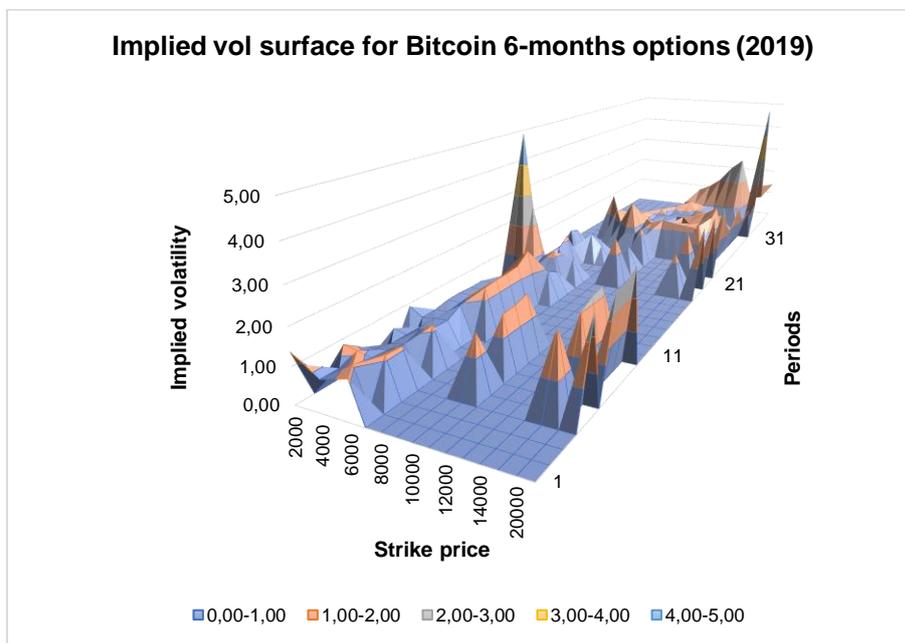

*Figure 10 - 3D surface of implied volatility for 6-month options during 2019*

5. **Conclusions**

The main purpose of this research was to present and compare different approaches for predicting future volatility in the Bitcoin market. We explored various methods of volatility calculation and forecasting, focusing on the last available quotation date (31/07/2019). Initially, we evaluated the historical daily volatility based on the price series to analyze its trend over time. The last value of volatility obtained indicated a relatively high level compared to previous ones, highlighting a volatile period in the market.

Subsequently, we calibrated several GARCH models to predict future volatility, including more traditional single-regime models. According to the information criteria we used, the GARCH (1,1) model performed best in this category. We also considered multiple-regime models designed to capture the unique distributional characteristics of Bitcoin, which tend to vary over time. Surprisingly, these models performed worse in terms of information criteria. However, a dual regime model with two sub-models represented by standard GARCHs (1,1) showed potential for outperforming the results obtained by the optimal single-regime model in terms of long-term volatility forecasts. This suggests the importance of considering the atypical characteristics in the distribution of cryptocurrency returns.

It's worth noting that a more complex model specification of the Markov-switching GARCH (MS-GARCH) with more regimes and higher lags could potentially yield better results than the standard GARCH. Additionally, we examined the coefficients obtained by other calibrated GARCH models to detect the presence of the inverse leverage effect in the Bitcoin series, as often identified in previous literature. However, these models did not produce consistent results. Nevertheless, the eGARCH model indicated the existence of a leverage effect, unlike the GJR-GARCH and apARCH models, which had insignificant leverage effect coefficients. As a result, this study cannot conclusively prove its hypothesis, and further in-depth analysis may be warranted. This could involve exploring different time periods or alternative models capable of detecting the existence of a leverage effect.

Regarding the implied volatility extracted from Bitcoin options listed on LedgerX and calculated with the Black-Scholes formula, it's important to note that the market was still not liquid enough to derive highly significant and representative volatility values that reflect the entire market's expectations. However, there appears to be a skewed volatility profile favoring options with lower strikes, specifically in-the-money (ITM) calls and out-of-the-money (OTM) puts, at least for options with a higher level of open interest and less extreme moneyness.

In summary, when analyzing the three different measures of volatility, including historical volatility, forecasted volatility, and implied volatility, they all indicate exceptionally high levels of expected fluctuations for the following year. This suggests a strong sense of uncertainty, particularly considering the distant maturity phase of the asset under consideration.

Nevertheless, the continuous increase in the number of option contracts may contribute to the reduction of volatility associated with Bitcoin investments in the future. This reduction could occur through the use of options as portfolio hedging instruments and due to their information content regarding the market's expectations.

This study confirms that the levels of implied volatility, while not directly comparable, are consistent with the high levels obtained using other methods. In some cases, implied volatility even exceeds 100% for sufficiently long time horizons.

Possible extensions of this research could involve analyzing volatility over different time periods, utilizing methodologies that assess out-of-sample goodness of fit by comparing predictions to realized values. Alternatively, exploring alternative forms of volatility modeling, including a broader spectrum of GARCH models—potentially incorporating jumps or structural breaks—or other models to calibrate the implied volatility surface. These models, though more complex, are common in financial markets due to their less rigid assumptions compared to the Black-Scholes model, examples of which include the Heston model or SABR.

Additionally, simulation techniques, such as Monte Carlo simulations, could be employed to incorporate various possible scenarios in the volatility calculation, potentially yielding less extreme results by smoothing individual simulation values. Moreover, investigating different timeframes for calibrating volatility models or considering alternative timing for quotations, such as focusing on short-term trading with intra-day data, could provide valuable insights.

Finally, it's essential to note that the results presented in this study are based on price series provided by LedgerX for both the underlying asset and the options. Using data from a different provider or trading platform may lead to significantly different results.

**References - Websites**